\documentclass{preprint}
\usepackage{times}
\usepackage{mhequ}
\usepackage{mhenvs}
\usepackage{mhfig}
\usepackage{array}
\usepackage{Symbols}
\usepackage{ScriptFonts}
\usepackage[ps2pdf,colorlinks,citecolor=blue,linkcolor=blue,urlcolor=blue]%
 {hyperref}

\makeatletter
\renewcommand{\theequation}{\arabic{section}.\arabic{equation}}
\@addtoreset{equation}{section}

\makeatother

\bgroup\theorembodyfont{\rmfamily}

\newtheorem{assumption}[lemma]{Assumption}
\egroup

\newenvironment{case}[1][r]{\left\{\begin{array}{#1>{\text\bgroup}l<{\egroup}}}{\end{array}\right.}

\def\XX{{\mathsf X}}

\def\PP{{\mathbf \Phi}}
\def\Par{{\cscr P}}
\def\BB{{\cscr B}}
\def\PR{\CP}
\def\OQ{{\mathsf P}}
\def\WA{{W_{\!L}}}

\def\rmu{\mu_0}
\def\={-\penalty-10000}

\def\Exp{{\symb E}}

\def\SEIGTH{{\textstyle{7\over 8}}}
\def\TQUAR{{\textstyle{3\over 4}}}
\def\HALF{{\textstyle{1\over 2}}}

\begin{document}

\title{Exponential Mixing for a Stochastic PDE \\ Driven by Degenerate Noise}
\author{M.~Hairer}
\institute{D\'epartement de Physique Th\'eorique, Universit\'e de Gen\`eve\\
\email{Martin.Hairer@physics.unige.ch}}
\date{March 27, 2001}

\maketitle\thispagestyle{empty}

\begin{abstract}
We study stochastic partial differential equations of the reaction-diffusion type. We show that, even if the forcing is very degenerate (\ie has not full rank), one has exponential convergence towards the invariant measure. The convergence takes place in the topology induced by a weighted variation norm and uses a kind of (uniform) Doeblin condition.
\end{abstract}

\section{Model and Result}

We consider the stochastic partial differential
equation given by
\begin{equ}[e:SGL]
du = \d_\xi^2 u\,dt - \OQ(u)\,dt + Q\,dW(t)\;,\qquad u \in W_{\rm per}^{(1,2)}([0,1])\;.
\tag{SGL}
\end{equ}
In this equation, $\OQ$ is a polynomial of odd degree with positive leading
coefficient and $\deg \OQ \ge 3$, $dW$ is the cylindrical
Wiener process on $\CH \equiv W_{\rm per}^{(1,2)}([0,1])$, and $Q:\CH \to \CH$ is a compact
operator which is diagonal in the trigonometric basis. The symbol $\xi \in [0,1]$ denotes the spatial variable. Further conditions on the spectrum of $Q$ will be made precise below.

In a recent paper \cite{EH3}, to which we also refer for further details about the model, it was shown that this equation possesses a unique invariant measure and satisfies the Strong Feller property. However, the question of the rate of convergence towards the invariant measure was left open. The aim of this paper is to show that this rate is exponential.

%This raised the question whether the convergence towards this invariant measure takes place exponentially fast. The answer to this question is positive and will be given in this paper.

There is a fair amount of very recent literature about closely related questions, mainly concerning ergodic properties of the 2D Navier-Stokes equation. To the author's knowledge, the main results are exposed in the works of Kuksin and Shirikyan \cite{KS00, ArmenKuk}, Bricmont, Kupiainen and Lefevere \cite{BKL,BKLExp}, and E, Mattingly and Sinai \cite{EMS,MatNS}, although the problem goes back to Flandoli and Maslowski \cite{FM}. The main differences between the model exposed here and the above papers is that we want to consider a situation where the {\em unstable} modes are {\em not} forced, whereas the forcing only acts onto the stable modes and is transmitted to the whole system through the nonlinearity. From this point of view, we are in a hypoelliptic situation where H\"ormander-type conditions apply \cite{H1,Ho}, as opposed to the essentially elliptic situation where the unstable modes are all forced and the (infinitely many) other modes are stabilized by the linear part of the equation.

 Returning to the model \eref{e:SGL}, we denote by $q_k$ the
 eigenvalue of $Q$ corresponding to the $k$th trigonometric function (ordered in such a way that $k > 0$). We make the following assumption on the $q_k$:

\begin{assumption}\label{ass:cond}
There exist constants
$k_* > 0$, $C_1 > 0$, $C_2>0$, $\alpha \ge 2$ and $\beta \in (\alpha - 1/8,\alpha]$ such that
\begin{equ}[e:cond]
C_1 k^{-2\alpha} \le q_k \le C_2k^{-2\beta} \;,\quad\text{for $k > k_*$.}
\end{equ}
There are no assumptions on $q_k$ for $k \le k_*$, in particular one may have $q_k=0$ in that region. Furthermore, $k_*$ can be chosen arbitrarily large.
\end{assumption}

We denote by $\Phi_t(u)$ the solution of \eref{e:SGL} at time $t$ with initial condition $u \in \CH$. If $\Phi_t$ exists and is sufficiently regular, one can define the semigroup $\CP^t$ acting on bounded functions $\phi$ and the semigroup $\CP_*^t$ acting on finite measures $\mu$ by
\begin{equ}
\bigl(\CP^t \phi\bigr)(u) = \Exp \Bigl(\bigl(\phi\circ\Phi_t\bigr)(u)\Bigr)\;,\qquad \bigl(\CP_*^t \mu\bigr)(A) = \Exp \Bigl(\bigl(\mu \circ \Phi_t^{-1}\bigr)(A)\Bigr)\;.
\end{equ}

In a recent paper \cite{EH3}, to which we also refer for further details about the model, it was shown that the above model satisfies the following.
\begin{theorem}\label{theo:SF}
Under Assumption~\ref{ass:cond}, the solution of \eref{e:SGL} defines a unique
stochastic flow $\Phi_t$ on $\CH$, thus also defining a Markov semigroup $\CP^t$.
The semigroup $\CP^t$ is Strong Feller and open set irreducible in arbitrarily short time. As a consequence, the semigroup $\CP_*^t$ acting on measures possesses a unique invariant measure on $\CH$.
\end{theorem}

Recall that a semigroup is said ``open set irreducible in arbitrarily short time'' if the probability of reaching a given open set in a given time is always strictly positive.

We denote by $\mu_*$ the unique invariant probability measure of \theo{theo:SF}.
We will show in this paper that for every probability measure $\mu$, we have $\CP_*^t\mu \to \mu_*$ and that this convergence takes place with an exponential rate (in time). More precisely, we introduce, for a given (possibly unbounded) Borel function $V : \CH \to [1,\infty]$, the {\em weighted variational norm} defined on every signed Borel measure $\mu$ by
\begin{equ}
\nnn{V}{\mu} \equiv \int_{\CH}V(x)\,\mu_+(dx) + \int_{\CH}V(x)\,\mu_-(dx)\;,
\end{equ}
where $\mu_\pm$ denotes the positive (resp.\ negative) part of $\mu$. When $V(x) = 1$, we recover the usual variational norm which we denote by $\nnn{}{\,\cdot\,}$.
We also introduce the family of norms $\|\cdot\|_\gamma$ on $\CH$ defined by
\begin{equ}
\|x\|_\gamma = \|L^\gamma x\|\;,
\end{equ}
where $L$ is the differential operator $1-\d_\xi^2$ and $\|\cdot\|$ is the usual norm on $\CH$, \ie
\begin{equ}
\|u\|^2 = \int_0^1 \bigl(|u|^2 + |\d_\xi u|^2\bigr)\,d\xi\;.
\end{equ}
The exact formulation of our convergence result is

\begin{theorem}\label{theo:main}
There exists a constant $\lambda > 0$ such that for every $p \ge 1$, every $\gamma \le \alpha$, and every probability measure $\mu$ on $\CH$, one has
\begin{equ}
\nnn{V_{\gamma,p}}{\CP_*^t\mu - \mu_*} \le C e^{-\lambda t}\;,\qquad \text{with}\;V_{\gamma,p}(u) = \|u\|_\gamma^p + 1\;,
\end{equ}
for every $t \ge 1$. The constant $C$ is independent of the probability measure $\mu$.
\end{theorem}

In the sequel, we will denote by $\PP$ the Markov chain obtained by sampling the solution of \eref{e:SGL} at integer times and by $\CP(x,\,\cdot\,)$ the corresponding transition probabilities.
\theo{theo:main} is a consequence of the following features of the model \eref{e:SGL}.
\begin{claim}
\item[\it A.] We construct a set $K$ having the property that there exists a 
probability measure $\nu$ and a constant $\delta > 0$ such that 
$\CP(x,\,\cdot\,) \ge \delta \, \nu(\,\cdot\,)$ for every $x\in K$. This means 
that $K$ behaves ``almost'' like an atom for the Markov chain $\PP$. This is 
shown to be a consequence of the Strong Feller property and the irreducibility of the Markov semigroup associated to \eref{e:SGL}.
\item[\it B.] The dynamics has very strong contraction properties in the sense that it reaches some compact set very quickly. In 
particular, one can bound uniformly from below the transition probabilities to a set $K$ satisfying property {\it A.}
\end{claim}

These conditions yield some strong Doeblin condition and thus lead to exponential convergence results.
The intuitive reason behind this is that, for any two initial measures, their image
under $\CP_*$ has a common part, the amount of which can be bounded uniformly from
below and cancels out. This will be clarified in the proof of \prop{prop:exp} below.

The remainder of the paper is organized as follows. In \sect{sec:PF}, we show 
how to obtain \theo{theo:main} from the above properties. The 
proof will be strongly reminiscent of the standard proof of the 
Perron-Frobenius theorem. In \sect{sec:contr} we then show the contraction 
properties of the dynamics and in \sect{sec:small} we show that every compact 
set has the property {\it A.}

\subsection*{Acknowledgements}

{\small
Several proofs in this paper are inspired by the recent work \cite{Luc} of Luc Rey-Bellet and Larry Thomas. The author is very grateful to them for having communicated their ideas before publication. The author also
benefitted from several useful discussions with Jean-Pierre Eckmann and Jacques Rougemont.
This work was partially supported by the Fonds National Suisse.}

\section{A Variant of the Perron-Frobenius Theorem}
\label{sec:PF}

The following proposition shows, reformulated in a more rigorous way, why the properties {\it A.}\ and {\it B.}\ yield exponential convergence results towards the invariant measure.

\begin{proposition}\label{prop:exp}
Let $\Psi$ be a Markov chain on a measurable space $\XX$ and let $\Psi$ satisfy the following properties:
\begin{claim}
\item[\it a.] There exist a measurable set $K$, a positive constant $\delta$ and a probability measure $\nu_*$ such that for every measurable set $A$ and every $x \in K$, one has $\CP(x,A) \ge \delta \,\nu_*(A)$.
\item[\it b.] There exists a constant $\delta' > 0$ such that $\CP(x,K) \ge \delta'$ for every $x\in\XX$.
\end{claim}
Then $\Psi$ has a unique invariant measure $\mu_*$ and one has for every 
probability measure $\mu$ the estimate $\nnn{}{\CP_*^n\mu - \mu_*} \le 2
(1-\delta\delta')^{-n/2}$. 
\end{proposition}

\begin{proof}
The first observation we make is that for every probability measure $\mu$ one has by property {\it a.},
\begin{equ}
  \bigl(\CP_*\mu\bigr)(K) = \int_\XX \CP(x,K)\,\mu(dx) \ge \delta'\;.
\end{equ}
As a consequence of this and of property {\it b.}, one has for every measurable set $A$ the bound
\begin{equ}[e:contr]
  \bigl(\CP_*^2\mu\bigr)(A) \ge \int_K \CP(x,A)\,\bigl(\CP_*\mu\bigr)(dx) \ge \delta\delta' \nu_*(A)\;.
\end{equ}
Define the constant $\eps = \delta\delta'$.
An immediate consequence of \eref{e:contr} is that for any probability measure $\mu$, one has
\begin{equ}
\nnn{}{\CP_*^2 \mu - \eps\nu_*} = 1-\eps\;.
\end{equ}
Now take any two probability measures $\mu$ and $\nu$. Denote by $\eta_\pm$ 
the positive (resp.\ negative) part of $\mu - \nu$. Since $\mu$ and $\nu$ are 
probability measures, one has $\nnn{}{\eta_+} = \nnn{}{\eta_-} = \Delta$, say. Then, since $\CP_*$ preserves probability, one has
\begin{equs}
\nnn{}{\CP_*^2 \mu - \CP_*^2 \nu} &= \nnn{}{\CP_*^2 \eta_+ - \CP_*^2 \eta_-} \le \nnn{}{\CP_*^2 \eta_+ - \Delta \eps\nu_*} 
 + \nnn{}{\CP_*^2 \eta_- - \Delta \eps\nu_*} \\
&\le 2\Delta(1-\eps) = (1-\eps)\nnn{}{\mu-\nu}\;.
\end{equs}
This completes the proof of \prop{prop:exp}.
\end{proof}

\theo{theo:main} is then an easy consequence of the following lemmas.

\begin{lemma}\label{lem:contract}
For every $\gamma \le \alpha$, every $t > 0$, and every $p \ge 1$, there exists a constant $C_{\gamma,p,t}$ such that for every finite measure $\mu$ on $\CH$ one has
\begin{equ}[e:condV]
\nnn{V_{\gamma,p}}{\CP_*^t\mu}\le C_{\gamma,p,t}\nnn{}{\mu}\;,
\end{equ}
with $\CP_*^t$ the semigroup acting on measures solving \eref{e:SGL}.
\end{lemma}

\begin{lemma}\label{lem:small}
For every compact set $K \subset \CH$, there exists a probability measure $\nu_*$ and a constant $\delta > 0$ such that $\CP(x,\,\cdot\,) \ge \delta\,\nu_*(\,\cdot\,)$ for every $x \in K$.
\end{lemma}

\begin{proof}[of \theo{theo:main}]
Fix once and for all $\gamma \le \alpha$ and $p \ge 1$. By \lem{lem:contract}, there exist constants $C$ and $\delta$ such that the set $K = \{x \in \CH\;|\; \|x\|_\gamma \le C\}$ satisfies $\CP(x,K) \ge \delta$ for every $x \in \CH$. By \lem{lem:small}, we can apply \prop{prop:exp} to find
\begin{equ}
\nnn{}{\CP_*^n \mu - \mu^*} \le 2e^{-\lambda n}\;,
\end{equ}
for some $\lambda > 0$ and for $n$ any integer. Since $\CP_*^t$ preserves positivity and probability, one immediately gets the same estimate for arbitrary real times. By \lem{lem:contract} and the invariance of $\mu_*$, this yields for some constant $C$,
\begin{equ}
\nnn{V_{\gamma,t}}{\CP_*^{t+1} \mu - \mu^*} \le C e^{-\lambda t}\;.
\end{equ}
The proof of \theo{theo:main} is complete.
\end{proof}
\begin{remark}
Writing $V$ instead of $V_{\gamma,p}$, condition \eref{e:condV} is equivalent to the statement that $\Exp_x V(\PP) \le C$ for all $x\in\CH$. It is also possible to achieve exponential convergence results if this condition is replaced by the weaker condition that
\begin{equ}[e:condV']
\Exp_x V(\PP) \le \begin{case}[c] c V(x) & for $x \in \CH\setminus K$,\\
\Lambda & for $x\in K$,\end{case}
\end{equ}
with $c \in (0,1)$, $\Lambda > 0$ and $K$ some compact set. The proof is somewhat lengthy and so we do not give it here. The interested reader is referred to \cite{MT,Luc}. The difference in the results is that one gets an estimate of the type
\begin{equ}
\nnn{V}{\CP_*^{t+1} \mu - \mu^*} \le C e^{-\lambda t}\nnn{V}{\mu}\;.
\end{equ}
So strong convergence towards the invariant measure holds for measures with finite $\nnn{V}{\,\cdot\,}$-norm and not necessarily for every probability measure.
\end{remark}
The remainder of the paper is devoted to the proof of Lemmas~\ref{lem:contract} and \ref{lem:small}.

\section{Contraction Properties of the Dynamics}
\label{sec:contr}

This section is devoted to the proof of \lem{lem:contract}. We reformulate it in a more convenient way as

\begin{proposition}\label{prop:bound}
For every $p \ge 1$, every $\gamma \le \alpha$, and every time $t>0$, there is a constant $C_{p,t,\gamma}>0$ such that, for every $x \in \CH$, one has
\begin{equ}[e:bound]
\Exp\bigl(\|\Phi_t(x)\|_\gamma^p\bigr) \le C_{p,t,\gamma}\;.
\end{equ}
\end{proposition}

\begin{proof}
We define the linear operator $L = 1-\d_\xi^2$ and the stochastic convolution
\begin{equ}
\WA(t) = \int_0^t e^{-L(t-s)}\,Q\,dW(s)\;.
\end{equ}
With these notations, the solution of \eref{e:SGL} reads
\begin{equ}[e:sol]
\Phi_t(x) = e^{-Lt}x + \int_0^t e^{-L(t-s)}\OQ\bigl(\Phi_s(x)\bigr)\,ds + \WA(t)\;.
\end{equ}
In a first step, we show that for every couple of times $0<t_1<t_2$, there exists a constant $C_{p,t_1,t_2}$ independent of the initial condition $x$ such that
\begin{equ}[e:sup]
\Exp\Bigl(\sup_{t_1 < s < t_2}\|\Phi_s(x)\|_\infty \Bigr) \le C_{p,t_1,t_2}\;.
\end{equ}
For this purpose, we introduce the auxiliary process $\Psi_t(x)$ defined by $\Psi_t(x) = \Phi_t(x) - \WA(t)$. We have for $\Psi_t$ the equation
\begin{equ}
\Psi_t(x) = e^{-Lt}x + \int_0^t e^{-L(t-s)}\OQ\bigl(\Psi_s(x)+\WA(s)\bigr)\,ds\;,
\end{equ}
\ie $\Psi_t(x)$ can be interpreted pathwise as the solution of the PDE
\begin{equ}[e:ord]
\dot \Psi_t = -L \Psi_t + \OQ\bigl(\Psi_t + \WA(t)\bigr)\;,\quad \Psi_0=x\;.
\end{equ}
If we denote by $q$ the degree of $\OQ$ (remember that $q \ge 3$), we have, thanks to the dissipativity of $L$, the inequality
\begin{equ}[e:ineq]
{D^- \|\Psi_t\|_\infty \over Dt} \le c_1 - c_2 \|\Psi_t\|_\infty^q + c_3 \|\WA(t)\|_\infty^q\;,
\end{equ}
where the $c_i$ are some strictly positive constants and $D^-/Dt$ denotes the left lower Dini derivative. An elementary computation allows to verify that the solutions of the ordinary differential equation $\dot y = -cy^q + f(t)$ (with positive initial condition and $f(s) > 0$) satisfy the inequality
\begin{equ}[e:solpol]
y(t) \le (qct)^{-1/(q-1)} + \int_0^tf(s)\,ds\;,
\end{equ}
independently of the initial condition. Standard estimates on Gaussian processes show furthermore that for every $t>0$ and every $p \ge 1$, there exists a constant $C_{p,t}$ such that
\begin{equ}
\Exp\bigl(\sup_{s \in [0,t]}\|\WA(s)\|_\infty^p\bigr)\le C_{p,t}\;.
\end{equ}
Combining this with \eref{e:solpol}, we get \eref{e:sup}.

It remains to exploit the dissipativity of the linear operator $L$ and the local boundedness of the nonlinearity to get the desired bound \eref{e:bound}. We write for $s \in [0,t/2]$ the solution of \eref{e:SGL} as
\begin{equ}
\Phi_{t/2+s}(x) = e^{-L(s+t/4)}\Phi_{t/4}(x) + \int_0^s e^{-L(s-r)}\OQ\bigl(\Phi_{{t/2}+r}(x)\bigr)\,dr + \WA(s)\;.
\end{equ}
Since $\|e^{-Lt}x\|\le t^{-1/2}\|x\|_\infty$, we have
\begin{equs}
\Exp\Bigl(\sup_{0<s<t/2} &\|\Phi_{t/2+s}(x)\|^p\Bigr) \le C_{p,t} + C_{p,t} \Exp\bigl(\sup_{0<s<t/2} \|\WA(s)\|^p\bigr) \\
&\quad +  C\Exp\biggl(\sup_{0<s<t/2}\Bigl(\int_0^s (s-r)^{-1/2}\bigl\|\OQ\bigl(\Phi_{{t/2}+r}(x)\bigr)\bigr\|_\infty\,dr\Bigr)^p\biggr) \\
&\le C_{p,t} + C_{p,t}\Exp\Bigl(\sup_{0<s<t} \bigl\|\Phi_{{t/2}+r}(x)\bigr\|_\infty^{pq}\Bigr) \le C_{p,t}\;,
\end{equs}
where we used the fact that
$
\Exp\bigl(\sup_{s \in [0,t]}\|\WA(s)\|_\gamma^p\bigr)
$
is finite for every $\gamma \le \alpha$, every $t>0$ and every $p\ge 1$. This technique can be iterated, using the fact that $\|e^{-Lt}x\|_{\gamma + 1/2} \le t^{-1/2}\|x\|_\gamma$, until one obtains the desired estimate \eref{e:bound}. The proof of \prop{prop:bound} is complete.
\end{proof}

\section{Strong Feller Chains and Small Sets}
\label{sec:small}

The aim of this section is to show that a sufficient condition for the existence of sets with the property {\it a.}\ of \prop{prop:exp} is that the Markov chain is open set irreducible and has the Strong Feller property.

We follow closely \cite{MT} in our definitions. The main difference with
their results is that we drop the assumption of local compactness of the
topological base space and that our estimates hold globally with respect to the initial condition. We will adopt the following notations:

The symbol $\XX$ stands for an arbitrary Polish space, \ie a complete, separable metric space. The symbol $\PP$ stands for a Markov chain on $\XX$. We denote by $\PR(x,A)$ the transition probabilities of $\PP$. The $m$-step transition probabilities are denoted by $\PR^m(x,A)$.
The symbol $\BB(\XX)$ stands for the Borel $\sigma$-field of $\XX$.

\begin{definition}
A set $K \in \BB(\XX)$ is called {\em small} if there exists an integer $m > 0$, a probability measure $\nu$ on $\XX$, and a constant $\delta > 0$ such that $\PR^m(x,A) \ge \delta \nu(A)$ for every $x \in K$ and every $A \in \BB(\XX)$. If we want to emphasize the value of $m$, we call a set {\em $m$-small}.
\end{definition}

With this definition, we reformulate \lem{lem:small} as

\begin{theorem}\label{theo:small}
If $\PP$ is irreducible and Strong Feller, every compact set is $2$-small.
\end{theorem}

The main step towards the proof of Theorem~\ref{theo:small} is to show the existence of small sets which are sufficiently big to be ``visible'' by the dynamics. Recall that a set $A$ is said to be {\em accessible} if $\CP(x,A) > 0$ for every $x \in \XX$. One has,

\begin{proposition}\label{prop:existsmall}
If $\PP$ is irreducible and Strong Feller, there exist accessible small sets.
\end{proposition}

\begin{proof}[of Theorem~\ref{theo:small}]
Recall that Doob's theorem guarantees the existence of a probability measure 
$\rmu$ such that the transition probabilities $\CP(x,\,\cdot\,)$ are all 
equivalent to $\rmu$. This is a consequence of the Strong Feller property and 
the irreducibility of $\PP$. 

By Proposition~\ref{prop:existsmall} there exists a small set $A$ such that $\rmu(A)>0$. For every $x \in \XX$ and every arbitrary $D \in \BB(\XX)$, we then have
\begin{equ}
\PR^{m+1}(x,D) \ge \int_{A}\PR(y,D)\,\PR^m(x,dy) \ge \PR(x,A)\inf_{y \in A}\PR(y,D) \ge \delta \PR(x,A) \nu(D)\;,
\end{equ}
for some $m>0$, $\delta>0$ and a probability measure $\nu$. Since, by the Strong Feller property, the function $x \mapsto \PR(x,A)$ is continuous and, by the accessibility of $A$, it is positive, there exists for every compact set $C \subset \XX$ a constant $\delta' > 0$ such that
\begin{equ}
\inf_{x \in C} \PR^{m+1}(x,D) \ge \delta' \nu(D)\;.
\end{equ}
The proof of Theorem~\ref{theo:small} is complete.
\end{proof}
%The existence of small sets with positive $\rmu$-measure is proven in \cite[Thm.~5.2.2]{MT}, using only the assumptions that $\PP$ is $\rmu$-irreducible (\ie $\rmu(A)>0 \Rightarrow \PR(x,A) > 0$) and that $\BB(\XX)$ is countably generated (which is a consequence of the separability of $\XX$).

The next subsection is devoted to the proof of Proposition~\ref{prop:existsmall}.

\subsection{Existence of accessible small sets}

In this subsection, we will work with partitions of $\XX$. We introduce the following notation: if $\Par$ is a partition of $\XX$, we denote by $\Par(x)$ the (only) element of $\Par$ that contains $x$. With this notation, one has the following theorem, a proof of which can be found \eg in \cite[p.~344]{Doob}.
\begin{theorem}[Basic Differentiation Theorem]\label{theo:diff}
Let $(\XX,{\cscr F},\mu)$ be a probability space and $\Par_n$ be an increasing sequence of finite measurable partitions of\/ $\XX$ such that the $\sigma$-field generated by $\bigcup_n \Par_n$ is equal to ${\cscr F}$. Let $\nu$ be a probability measure on $\XX$ which is absolutely continuous with respect to $\mu$ with density function $h$. Define the sequence of functions $h_n$ by
\begin{equ}
h_n(x) = \begin{case}[c] \displaystyle{\nu(\Par_n(x))\over \mu(\Par_n(x))} & if $\mu(\Par_n(x)) > 0$, \\[4mm]
0 & if $\mu(\Par_n(x)) = 0$.\end{case}
\end{equ}
Then there exists a set $N$ with $\mu(N) = 0$ such that $\lim_{n \to \infty} h_n(x) = h(x)$ for every $x \in \XX \setminus N$.
\end{theorem}
This theorem is the main ingredient for the proof of \prop{prop:existsmall}.

The first point one notices is that if $\XX$ is a Polish space, one can explicitly construct a sequence $\Par_n$ of partitions that generate the Borel $\sigma$-field. Choose a sequence $\{x_i\}_{i=1}^\infty$ of 
elements which are dense in $\XX$ (the existence of such a sequence is guaranteed by the separability of $\XX$) and a sequence $\{\eps_j\}_{j=1}^\infty$ 
such that $\eps_j > 0$ and $\lim_{j \to \infty}\eps_j = 0$. Denote by 
$\CB(x,r)$ the open ball of radius $r$ and center $x$.
We then define the sets $M_i^j$ ($i\ge 1$ and $j\ge 0$) by
\begin{equ}
M_i^0 = \XX\;, \quad
M_i^j = \CB(x_i,\eps_j)\;.
\end{equ}
This defines an increasing sequence of finite partitions $\Par_n$ by
$
\Par_n = \bigvee_{i,j\le n} \{ M_i^j \}
$
($\vee$ denotes the refinement of partitions).
We denote by ${\cscr F}_\infty$ the $\sigma$-field generated by $\bigcup_n \Par_n$. Since every open set $S \subset \XX$ can be written as a countable union
\begin{equ}
S = \bigcup \bigl\{M_i^j\;|\; M_i^j \subset S\bigr\}\;,
\end{equ}
the open sets belong to ${\cscr F}_\infty$ and so ${\cscr F}_\infty = \BB(\XX)$. This construction guarantees the applicability of the Basic Differentiation Theorem to our situation. We are now ready to give the

\begin{proof}[of \prop{prop:existsmall}]
Let us denote by $p(x,y)$ a jointly measurable 
version of the densities of $\CP(x,\,\cdot\,)$ with respect to $\rmu$.

We define for every $x,y \in \XX$ the sets $S_x \in \BB(\XX)$ and $S_y^* \in \BB(\XX)$ by
\begin{equ}
S_x = \bigl\{y\in\XX\,|\,p(x,y) > \HALF\bigr\}\;,\quad
S^*_y = \bigl\{x\in\XX\,|\,p(x,y) > \HALF\bigr\}\;,
\end{equ}
and the set $S^2 \in \BB(\XX\times\XX)$ by
\begin{equ}
S^2 = \bigl\{(x,y)\in\XX\times\XX\,|\,p(x,y) > \HALF\bigr\}\;.
\end{equ}
Since $\CP(x,\XX) = 1$ for every $x\in\XX$, one has $\rmu(S_x) > 0$ for every $x$ and therefore
$\rmu^2(S^2) = \int_\XX \rmu(S_x)\,d\rmu(x) > 0$, where $\rmu^2 = \rmu\times\rmu$. Define the subset $S^3$ of $\XX^3$ by
\begin{equ}[e:S3]
S^3 = \bigl\{(x,y,z)\in\XX^3\;|\;(x,y)\in S^2\;\text{and}\;(y,z) \in S^2\bigr\}\;.
\end{equ}
One has similarly $\rmu^3(S^3) = \int_{S^2} \rmu(S_y)\,d\rmu^2(x,y) > 0$. Let us now define the sets $\Par_n(x)$ as above and define $\Par_n(x,y) = \Par_n(x) \times \Par_n(y)$.

\begin{figure}
\begin{center}
\mhpastefig[2/3]{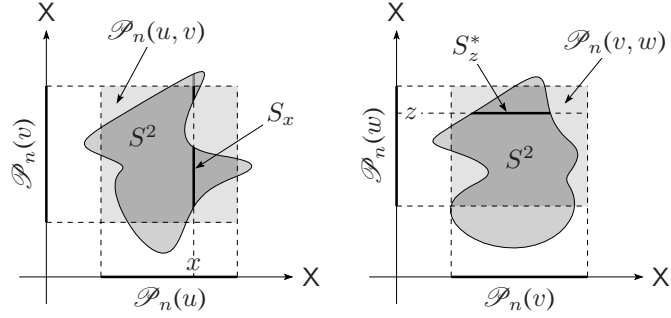}
\end{center}
\abovecaptionskip 0pt
\caption{Construction of $D$ and $E$.}\label{fig:sect}
\end{figure}
By \theo{theo:diff} with $\mu = \rmu^2$ and $\nu = \rmu^2 |_{S^2}$, there exists a $\rmu^2$-null set $N$ such that for $(x,y) \in S^2 \setminus N$ one has
\begin{equ}
\lim_{n\to \infty}{\rmu^2\bigl(S^2 \cap \Par_n(x,y)\bigr)\over \rmu^2\bigl(\Par_n(x,y)\bigr)} = 1\;.
\end{equ}
Since on the other hand $\rmu^3(S^3)>0$, there exist a triple $(u,v,w)$ and an integer $n$ such that $\rmu^2\bigl(\Par_n(u,v)\bigr) > 0$, $\rmu^2\bigl(\Par_n(v,w)\bigr) > 0$, and \minilab{e:covers}
\begin{equs}
\rmu^2\bigl(S^2 \cap \Par_n(u,v)\bigr) &\ge \SEIGTH\rmu^2\bigl(\Par_n(u,v)\bigr)\;,\label{e:covers1}\\
\rmu^2\bigl(S^2 \cap \Par_n(v,w)\bigr) &\ge \SEIGTH\rmu^2\bigl(\Par_n(v,w)\bigr)\;.\label{e:covers2}
\end{equs}
This means that $S^2$ covers simultaneously seven eights of the ``surfaces'' of both sets $\Par_n(u,v)$ and $\Par_n(v,w)$. (See Figure~\ref{fig:sect} for an illustration of this construction.) As a consequence of \eref{e:covers1}, the set
\begin{equ}
D = \bigl\{ x \in \Par_n(u)\;|\; \rmu\bigl(S_x \cap \Par_n(v)\bigr) \ge \TQUAR\rmu\bigl(\Par_n(v)\bigr)\bigr\}\;,
\end{equ}
satisfies $\rmu(D) \ge \HALF \rmu\bigl(\Par_n(u)\bigr)$. Similarly, the set
\begin{equ}
E = \bigl\{z \in \Par_n(w)\;|\; \rmu\bigl(S_z^* \cap \Par_n(v)\bigr) \ge \TQUAR\rmu\bigl(\Par_n(v)\bigr)\bigr\}\;,
\end{equ}
satisfies $\rmu(E) \ge \HALF \rmu\bigl(\Par_n(w)\bigr)$. On the other hand, one has by the definitions of $E$ and $D$ that for $x \in D$ and $z\in E$, $\rmu(S_x \cap S_z^*) \ge \HALF \rmu\bigl(\Par_n(v)\bigr)$. Thus
\begin{equ}[e:est]
p^2(x,z) \ge \int_{S_x \cap S_z^*}p(x,y)p(y,z)\,\rmu(dy) \ge {1\over 4}\rmu(S_x \cap S_z^*) \ge {1\over 8} \rmu\bigl(\Par_n(v)\bigr)\;,
\end{equ}
for every $x\in D$ and every $y\in E$.
Defining a probability measure $\nu$ by setting $\nu(\Gamma) = \rmu(\Gamma\cap E)/\rmu(E)$, there exists $\delta > 0$ such that for every $x\in D$, one has $\CP(x,\Gamma) \ge \delta \nu(\Gamma)$ and thus $D$ is small. Since $\rmu(D) > 0$, the proof of \prop{prop:existsmall} is complete.
\end{proof}

\bibliographystyle{myalph}
\markboth{\sc \refname}{\sc \refname}
\bibliography{refs}
\end{document}